\documentclass[fleqn,usenatbib]{mnras}

\usepackage[T1]{fontenc}

\DeclareRobustCommand{\VAN}[3]{#2}
\let\VANthebibliography\thebibliography
\def\thebibliography{\DeclareRobustCommand{\VAN}[3]{##3}\VANthebibliography}

\usepackage{graphicx}	
\usepackage{amsmath}	
\usepackage{amssymb}	

\usepackage{graphicx}
\newcommand\KRLow{KR03}
\newcommand\KRMed{KR06}
\newcommand\KRHigh{KR12}

\usepackage{newtxtext,newtxmath}
\title[Evolution of rotating star clusters]{Long-term evolution of multimass rotating star clusters}
\author[A. R. Livernois et al.]{Alexander R. Livernois,$^{1}$\thanks{E-mail: allivern@iu.edu}
Enrico Vesperini,$^{1}$
Anna Lisa Varri,$^{2,3}$
\newauthor
Jongsuk Hong,$^{4}$
Maria Tiongco$^{5}$
\\
$^{1}$Department of Astronomy, Indiana University, Bloomington, IN, 47405, USA\\
$^{2}$Institute for Astronomy, University of Edinburgh, Royal Observatory, Blackford Hill, Edinburgh EH9 3HJ, UK\\
$^{3}$School of Mathematics and Maxwell Institute for Mathematical Sciences, University of Edinburgh, Kings Buildings, Edinburgh EH9 3FD, UK\\
$^{4}$Korea Astronomy and Space Science Institute, 776 Daedeok-daero, Yuseong-gu, Daejeon 34055, Republic of Korea\\
$^{5}$University of Colorado, JILA and Department of Astrophysical and Planetary Sciences, 440 UCB, Boulder, CO 80309- USA\\
}

\begin{document}

\label{firstpage}
\pagerange{\pageref{firstpage}--\pageref{lastpage}}
\maketitle

\begin{abstract}
We investigate the long-term dynamical evolution of the internal kinematics of multimass rotating star clusters.
We have performed a set of N-body simulations to follow the internal evolution of clusters with different degrees of initial rotation and have explored the evolution of the rotational velocity, the degree of energy equipartition, and anisotropy in the velocity distribution. Our simulations show that: 1) as the cluster evolves, the rotational velocity develops a dependence on the stellar mass with more massive stars characterised by a more rapid rotation and a peak in the rotation curve closer to the cluster centre than low-mass stars; 2) the degree of energy equipartition in the cluster's intermediate and outer regions depends on the component of the velocity dispersion measured; for more rapidly rotating clusters, the evolution towards energy equipartition is more rapid in the direction of the rotational velocity; 3) the anisotropy in the velocity distribution is stronger for massive stars; 4) both the degree of mass segregation and energy equipartition are characterised by spatial anisotropy; they have a dependence on both $R$ and $z$, correlated with the flattening in the spatial variation of the cluster's density and velocity dispersion, as shown by 2D maps of the mass segregation and energy equipartition on the ($R$-$z$) meridional plane.

\end{abstract}


\section{Introduction}
A number of recent observational studies have significantly expanded the dynamical picture of globular clusters by including information about their internal kinematics and showing that these systems, traditionally considered non-rotating and isotropic, are often characterised by more complex properties including the presence of internal rotation and an anisotropic velocity distribution (see e.g. \citealt{2014FaMa}, \citealt{2017BeBi}, \citealt{2018FeLa}, \citealt{2018KaHu}, \citealt{2018LaFe}, \citealt{2019JiWe}, \citealt{2019SoBa}, \citealt{2021VaBa}).

Significant progress has been made also in the study of the structure, morphology, and rotation of young star clusters (see e.g. the results of the MYStIX project: \citealt{2014mystix}, \citeyear{2015mystixa}, \citeyear{2015mystixb},  \citeyear{2017mystix}, \citealt{2018mystix}; and \citealt{2012HeGi} from the Tarantula Survey) and data from  the {\it Gaia} mission have  allowed for richer insight into the dynamics of young clusters by adding information about their internal kinematic properties (see e.g.  \citealt{2019GeFe}, \citealt{2019KuHi}, \citealt{2020LiHo}, \citealt{2021DaVa}, \citealt{2021LiNa}, \citealt{2021SwDO}). 

On the theoretical side, significant efforts have been devoted to the study of the physical origin and development of dynamical properties such as internal rotation and velocity anisotropy, as well as their impact on established processes such as mass segregation during the star clusters' formation and very early evolutionary phases (see e.g. \citealt{2004GoWh}, \citealt{2007McVe}, \citealt{2009AlGo}, \citeyear{2010AlGo}, \citealt{2009MoBo}, \citealt{2012FuSa}, \citealt{2014VeVa}, \citealt{2014BaKr}, \citeyear{2017BaKr}, \citeyear{2018BaKr},  \citealt{2016FuPo}, \citealt{2016PaGo}, \citealt{2017DoFe}, \citealt{2017Ma},
\citealt{2018SiRi}, \citealt{2020DaPa}, \citealt{2020BaMa}, \citealt{2020LaNa}, \citealt{2021BaTo}, \citealt{2021LiVe}).

In particular, the recent observations showing the presence of internal rotation in many old Galactic globular clusters have spurred a renewed interest in the theoretical study of the dynamical evolution of rotating stars clusters. A number of investigations (see e.g. \citealt{1999EiSp}, \citealt{2004KiLe},  \citealt{2007ErGl}, \citealt{2013HoKi}, \citealt{2017TiVe}) have shown that, as an initially rotating cluster evolves, the outward transport of angular momentum and the loss of angular momentum carried away by escaping stars cause the rotational velocity to decrease thus suggesting that the present-day rotation observed in globular clusters is a lower limit on the cluster's initial rotation. These results have been supported by observational findings of a negative correlation between rotation speed and dynamical age of star clusters (see e.g. \citealt{2018Biva}, \citealt{2018KaHu}).

As observational studies continue to shed light on the internal dynamical properties of star clusters, new efforts aimed at expanding the theoretical framework necessary to interpret the observational results are necessary.
Many fundamental questions concerning the structural and kinematic evolution of rotating star clusters and the additional dynamical complexities introduced by the presence of rotation require further investigation (see e.g. \citealt{2013HoKi}, \citealt{2021BrRo} for the possible effects associated to the development of a bar in rapidly rotating systems, \citealt{2019SzMe}, \citealt{2021TiVa}, \citealt{2021LiVe} for the development of anisotropic mass segregation during the cluster's early and long-term evolution and its connection with the effects of vector resonant relaxation discussed by \citealt{2019MeKo}).

In this paper, we present the results of a set of N-body simulations of rotating clusters with different degrees of initial rotation, a spectrum of masses, and the effects of an external tidal field. The goal of our study is to explore the  long-term evolution of the clusters' kinematic properties and their dependence on the stellar mass, as well as the effects of rotation on the evolution towards energy equipartition and mass segregation.

The paper proceeds as follows. Section \ref{Methods} describes the N-body models; our results are presented in Section \ref{Results}, and we summarise our conclusions in Section \ref{Conclusions}.

\section{Methods and Initial Conditions}

\label{Methods}
Our study is based on the results of a set of three N-body simulations run with the GPU-accelerated version of the \textsc{NBODY6} code (\citealt{nbody6_2003}, \citealt{GPU_nbody6}) on the Indiana University's Carbonate supercomputer.
Our models have initially $N = 5\times 10^4$ stars with masses distributed according to a \cite{KIMF} initial mass function between 0.1 and 1 $M_{\sun}$. No primordial binaries are included in our initial conditions and, since we are focusing our attention on the effects of two-body relaxation dynamics, we do not include the effects of stellar evolution and the mass spectrum considered is that typical of an old star cluster.
The spatial and kinematic properties of our models are set using the distribution function of  rotating King models (see \citealt{1987LuGu}) with central dimensionless potential, $W_0=6$ and various degrees of internal rotation. Rotation is quantified by the dimensionless parameter  $\omega_0=3/(\sqrt{4\pi G \rho_{\rm c}})\Omega_{\rm c}$, where $G$ is the gravitational constant, $\rho_{\rm c}$ is the central mass density, and $\Omega_{\rm c}$ is the central angular velocity of the system. 
The models used here have $\omega_0=$ 0.3, 0.6, and 1.2 (see e.g. \citealt{2013HoKi} for a study of the evolution and stability of models with these parameters and a two-component mass distribution), and will hereafter be referred to as \KRLow, \KRMed, and \KRHigh, respectively. 

We have added a tidal field from the host galaxy, modelled as a point mass, and assumed circular orbits for our clusters. The clusters have a Jacobi radius, $r_{\rm J}$, initially equal to twice their initial total radius, $r_{\rm tot}$; stars moving beyond a radius equal to twice the Jacobi radius are removed from the simulation.

The simulations follow the evolution of the clusters until they reach core collapse, at about 3 $t_{\rm rh}$, where $t_{\rm rh}$ is the initial half-mass relaxation time of each cluster, defined, in N-body units, as:

\begin{equation}
    t_{\rm rh}=\frac{0.138 N r_{\rm h}^{3/2}}{\ln(0.02N)}
\end{equation}

\noindent 
where $r_{\rm h}$ and $N$ are, respectively, the initial 3D half-mass spherical radius and the initial number of stars.

\begin{figure*}
    \centering
    \includegraphics[width=0.9\textwidth]{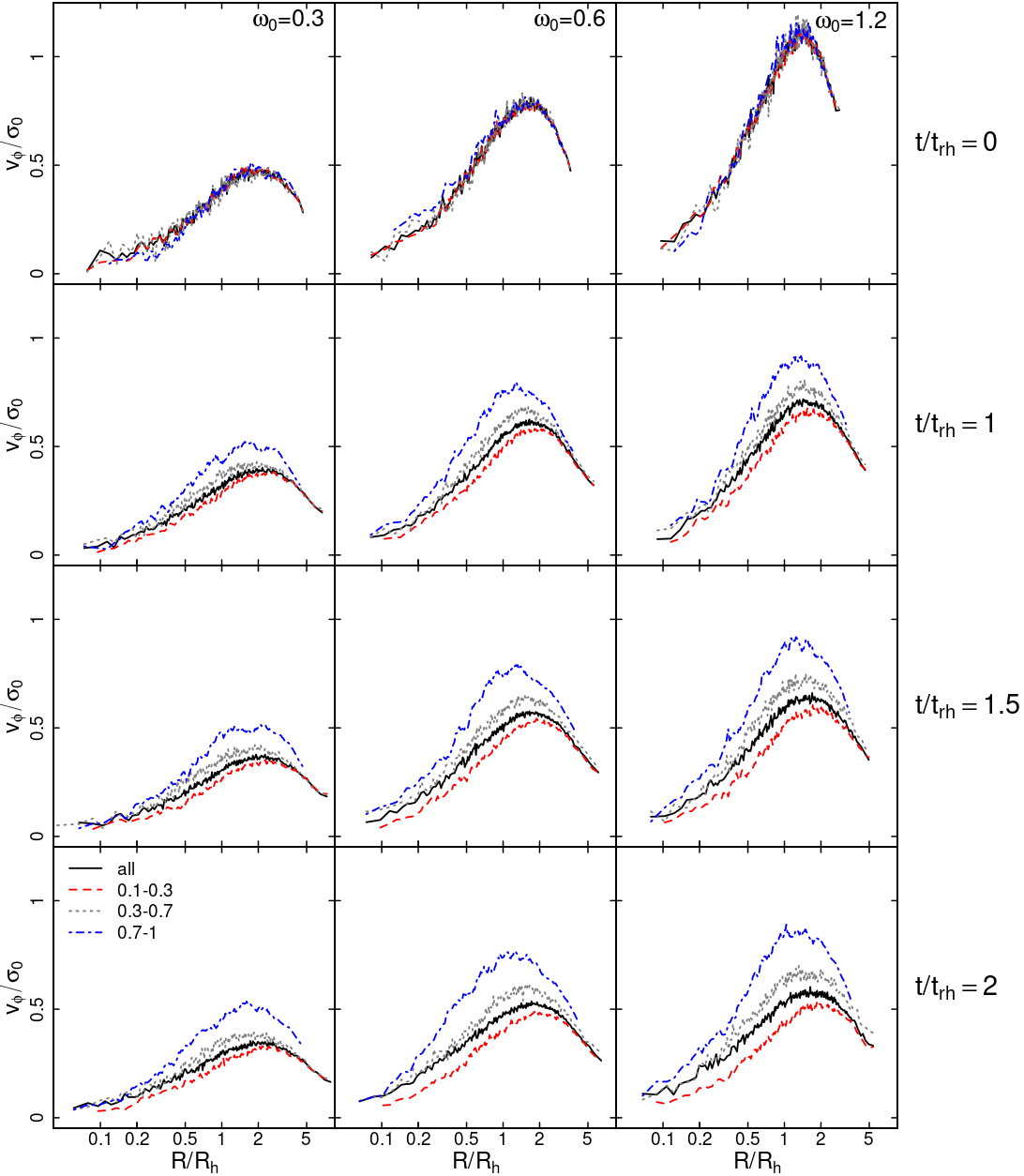}
    \caption{Time evolution of the radial profile of the rotation velocity, $v_{\rm \phi}$,  for each mass range in the legend (in units of $M_{\sun}$), normalised by the current central velocity dispersion of all stars, evaluated on the plane perpendicular to the rotation axis, $\sigma_{\rm 0}$, for the \KRLow\ (left), \KRMed\ (middle), and \KRHigh\ (right) models. The rotational velocity curves shown in this figure are calculated by combining snapshots over the time intervals 0-0.05, 0.9-1.1, 1.4-1.6, 1.9-2.1 $t_{\rm rh}$. $R$ and $R_{\rm h}$ are, respectively, the cylindrical radius and cylindrical half-mass radius. As the clusters evolve, their rotational velocities develop a dependence on the stellar mass.
}
    \label{fig:VC_disp}
\end{figure*}

\begin{figure*}
    \centering
    \includegraphics[width=\textwidth]{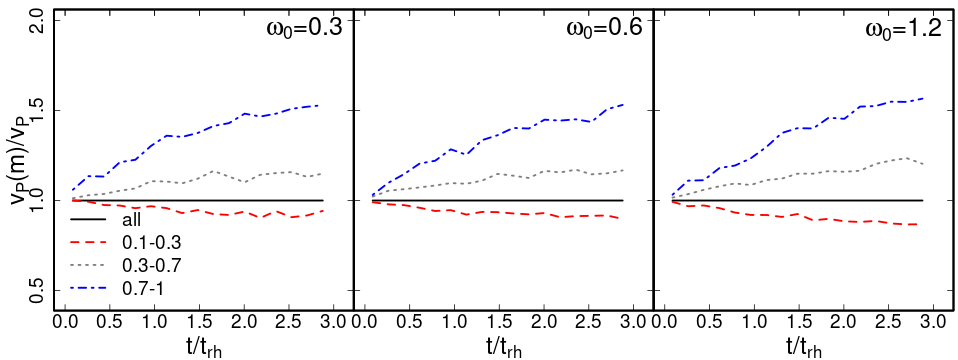}
    \caption{Time evolution of the peak velocity, $v_{\rm P}(m)$, of each mass range in the legend (in units of $M_{\sun}$), normalised by the peak velocity of all stars, $v_{\rm P}$, for the \KRLow\ (left), \KRMed\ (middle), and \KRHigh\ (right) models. All models develop a clear trend between stellar mass and peak velocity.}
    \label{fig:VPEAK}
\end{figure*}
\begin{figure*}
    \centering
    \includegraphics[width=\textwidth]{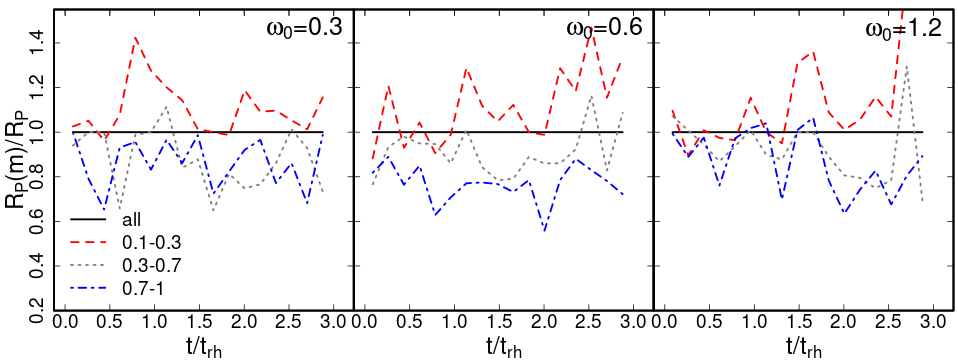}
    \caption{Time evolution of the cylindrical radius associated with $v_{\rm P}$, $R_{\rm P}$, of each mass range in the legend (in units of $M_{\sun}$), normalised by $R_{\rm P}$ of the entire cluster, for the \KRLow\ (left), \KRMed\ (middle), and \KRHigh\ (right) models. All models develop a trend between stellar mass and $R_{\rm P}$. }
    \label{fig:RPEAK}
\end{figure*}

\begin{figure*}
    \centering
    \includegraphics[width=\textwidth]{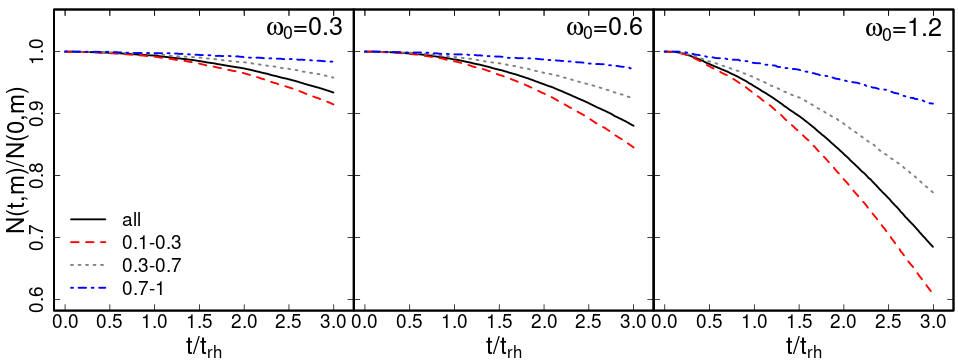}
    \caption{Time evolution of the total number of stars in each mass range in the legend (in units of $M_{\sun}$) remaining within one tidal radius of the star cluster, normalised by the initial number of stars in each mass range at $t=0$ for the \KRLow\ (left), \KRMed\ (middle), and \KRHigh\ (right) models.}
    \label{fig:mloss}
\end{figure*}

\begin{figure*}
    \centering
    \includegraphics[width=\textwidth]{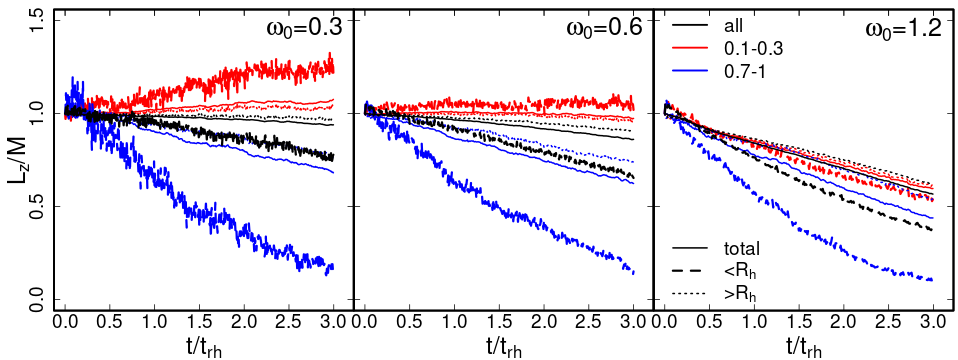}
    \caption{Time evolution of the total $z$ component of the angular momentum divided by the total mass ($L_{\rm z}/M$), normalised by the initial value of this ratio for each radial region and mass group in the legend, for the \KRLow\ (left), \KRMed\ (middle), and \KRHigh\ (right) models.}
    \label{fig:lzfrac}
\end{figure*}
\begin{figure*}
    \centering
    \includegraphics[width=0.9\textwidth]{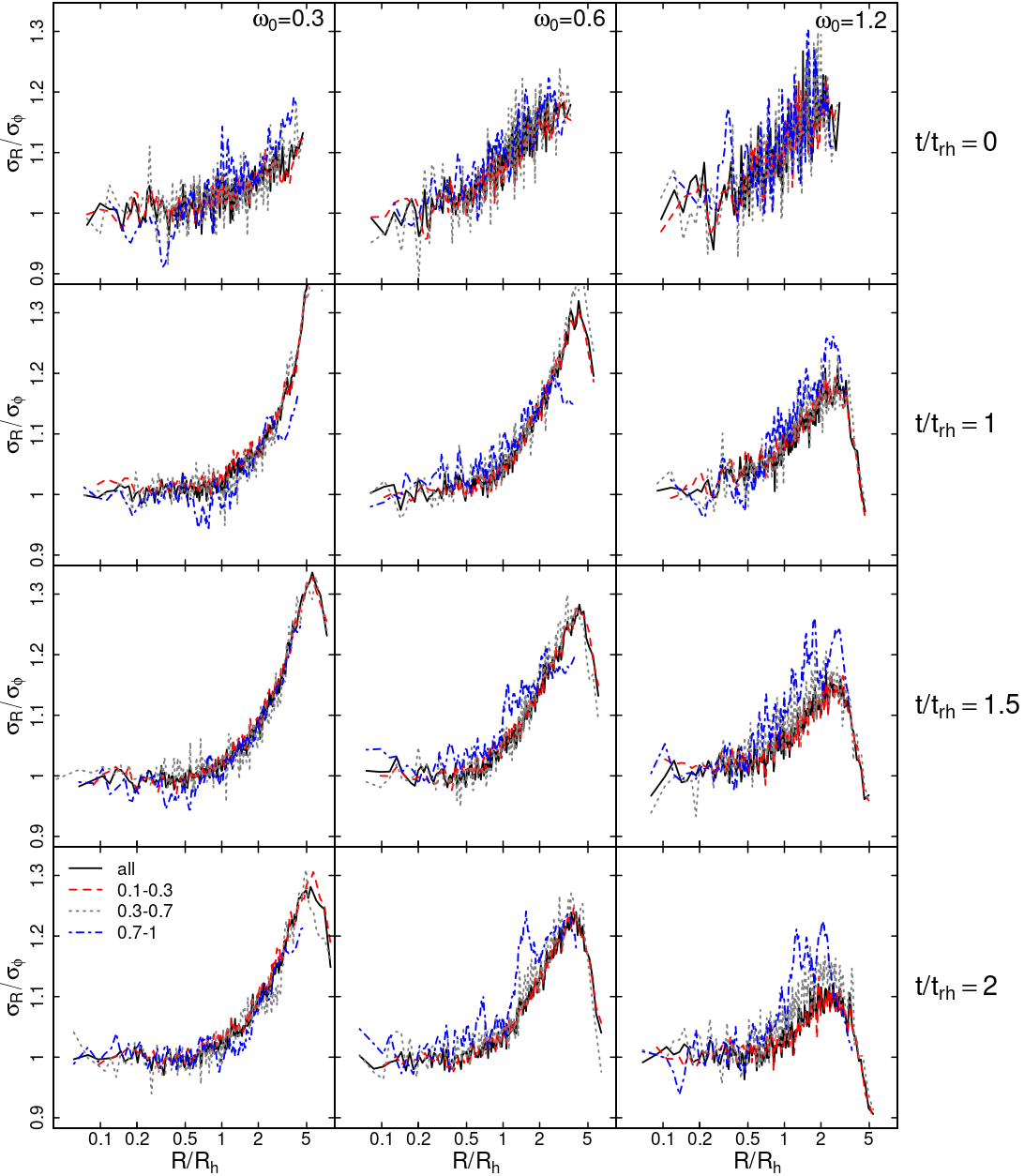}
    \caption{Time evolution of the radial profile of the velocity anisotropy, $\sigma_{\rm R}/\sigma_{\rm \phi}$, for each mass range in the legend (in units of $M_{\sun}$), for the \KRLow\ (left), \KRMed\ (middle), and \KRHigh\ (right) models. The anisotropy profiles shown in this figure are calculated by combining snapshots over the time intervals 0-0.05, 0.9-1.1, 1.4-1.6, 1.9-2.1 $t_{\rm rh}$. $R$ and $R_{\rm h}$ are, respectively, the cylindrical radius and cylindrical half-mass radius.}
    \label{fig:anisotropy}
\end{figure*}

\begin{figure*}
    \centering
    \includegraphics[width=\textwidth]{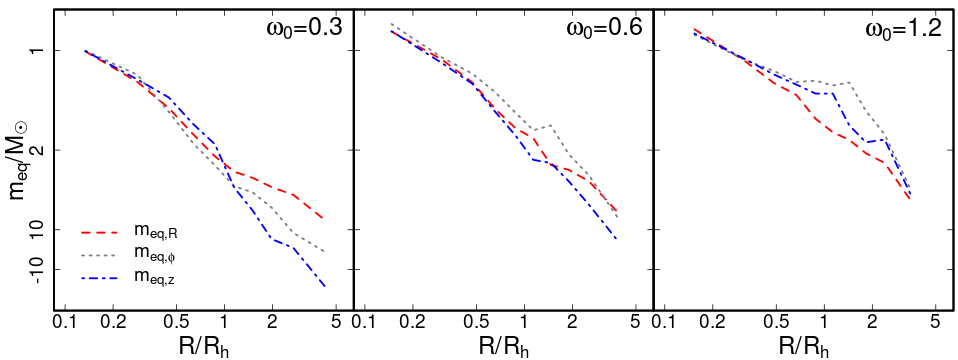}
    \caption{Radial profile of the equipartition mass, $m_{\rm eq}$ (in units of $M_{\sun}$, defined in equation \ref{eq:meq}), evaluated using each velocity component in the legend for the \KRLow\ (left), \KRMed\ (middle), and \KRHigh\ (right) models from $1.9 t_{\rm rh}<t<2.1 t_{\rm rh}$. The $y$-axis is in a reciprocal scale to clearly visualise the trends of $m_{\rm eq}$ and its inversion in the outer regions in the \KRLow\ model.}
    \label{fig:meq_inv}
\end{figure*}

\begin{figure*}
    \centering
    \includegraphics[width=\textwidth]{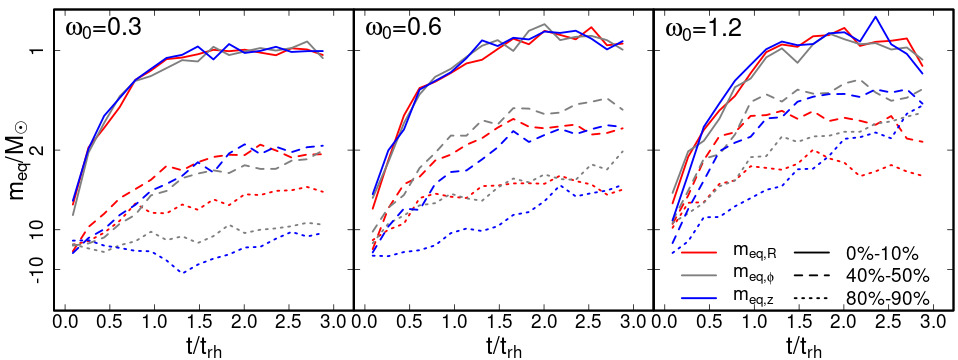}
    \caption{Time evolution of the equipartition mass, $m_{\rm eq}$ (in units of $M_{\sun}$, defined in equation \ref{eq:meq})  for the \KRLow\ (left), \KRMed\ (middle), and \KRHigh\ (right) models. $m_{\rm eq}$ is evaluated in different Lagrangian radial shells and using different velocity components as noted in the legend. The $y$-axis is in a reciprocal scale to clearly visualise the trends of $m_{\rm eq}$ and its inversion in the outer regions in the \KRLow\ model.}
    \label{fig:meq_time}
\end{figure*}

\begin{figure*}
    \centering
    \includegraphics[width=\textwidth]{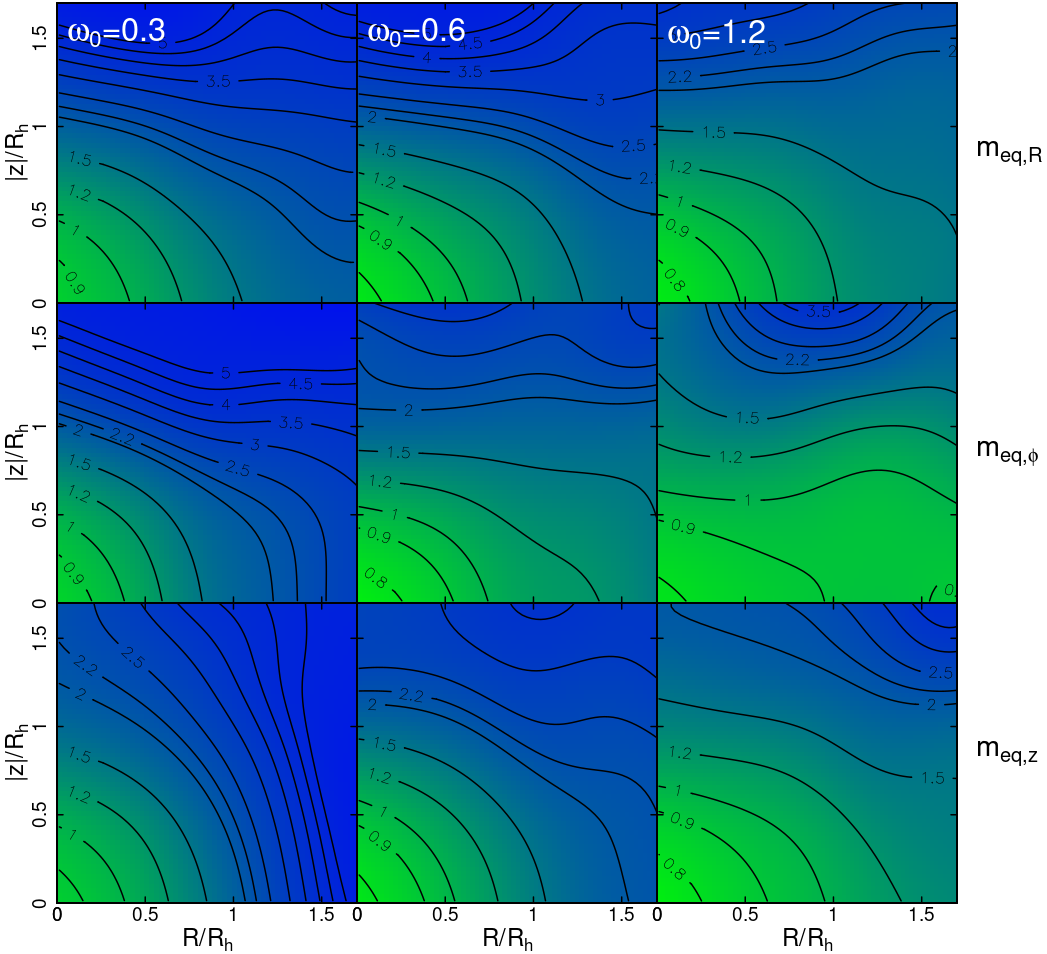}
    \caption{Spatial map of the equipartition mass  ($m_{\rm eq,R}$ (top), $m_{\rm eq,\phi}$ (middle), and $m_{\rm eq,z}$ (bottom); in units of $M_{\sun}$, defined in equation \ref{eq:meq}) in the $R$-$|z|$ plane for the \KRLow\ (left), \KRMed\ (middle), and \KRHigh\ (right) models at $2 t_{\rm rh}$.}
    \label{fig:meqmap}
\end{figure*}

\begin{figure*}
    \centering
    \includegraphics[width=\textwidth]{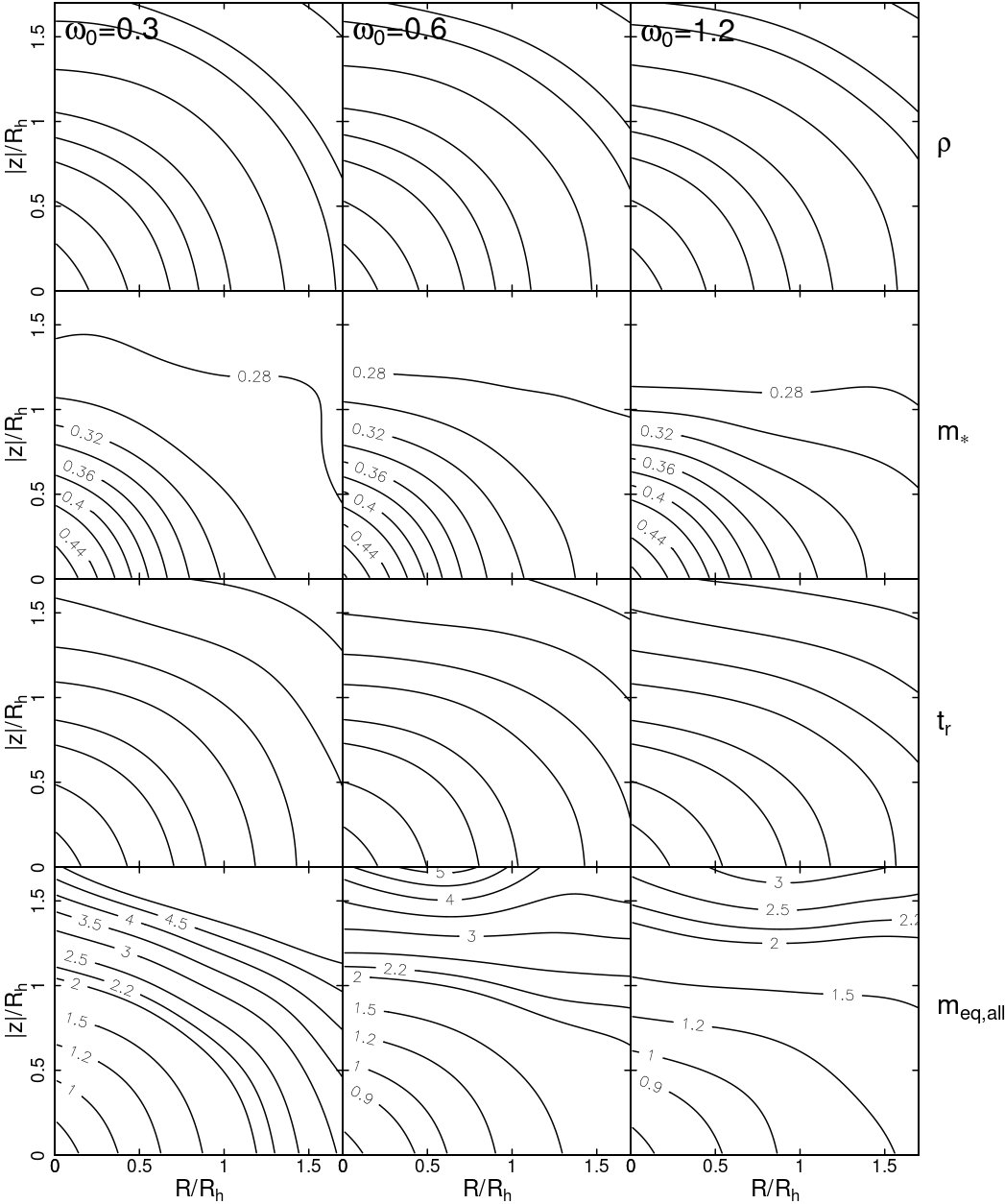}
    \caption{Spatial contours of the local mass density ($\rho$),  stellar mass ($m_*$, in solar units), $\sigma^3/(\left< m \right>\rho)$ (a quantity proportional to the relaxation time $t_{\rm r}$), and equipartition mass calculated using all velocities ($m_{\rm eq, all}$; in units of $M_{\sun}$, defined in equation \ref{eq:meq}), mapped in the $R$-$|z|$ plane for the \KRLow\ (left), \KRMed\ (middle), and \KRHigh\ (right) models at $2 t_{\rm rh}$. All of the models are characterised by a flattening of the contours along the $R$ direction, showing that in rotating models these quantities are functions of both $R$ and $z$.}
    \label{fig:LDMEQmap}
\end{figure*}

\section{Results}
\label{Results}
\subsection{Evolution of rotational profiles}
\label{sec:Rotation}
In this section, we start our analysis by investigating the evolution of the rotational properties of our models.

In Fig. \ref{fig:VC_disp}, we show the rotation curves for all models at times 0, 1, 1.5, and 2 $t_{\rm rh}$. The rotation curves are evaluated by calculating the median rotational velocity, $v_{\rm \phi}$, in concentric cylindrical radial bins aligned along the rotation axis. The rotational velocity is calculated both for all stars and for stars in different mass groups. 
These rotation curves are normalised to the current central velocity dispersion of all stars, evaluated on the plane perpendicular to the rotation axis: $\sigma_0= \sqrt{(\sigma^2_{\rm R}+\sigma^2_{\rm \phi})/2}$, where $\sigma_{\rm R}$ and $\sigma_{\rm \phi}$ are, respectively, the radial and tangential velocity dispersions on that plane. 

The rotational curve measured for all stars shows a general decrease similar to that found in previous studies (see e.g \citealt{1999EiSp}, \citealt{2004KiLe}, \citealt{2007ErGl}, \citealt{2013HoKi}, \citealt{2017TiVe}), which is associated with the loss of angular momentum carried away by escaping stars. The rotational curves measured for stars in different mass groups, on the other hand, reveal various trends resulting from the complex combination of the effects of  mass segregation, angular momentum transfer between different mass groups, and angular momentum loss.
Specifically, our simulations show the gradual development of a dependence of the rotational properties on the stellar mass: more massive stars tend to rotate more rapidly than low-mass stars and have the peak in their rotational velocity closer to the cluster centre than low-mass stars. 

Figs. \ref{fig:VPEAK} and \ref{fig:RPEAK} show the time evolution of the peak of $v_{\rm \phi}$, $v_{\rm P}$, and its distance from the cluster rotation axis, $R_{\rm P}$. 
These figures clearly illustrate that, as each cluster evolves, both of these quantities develop a dependence on the stellar mass.
\setcitestyle{notesep={; }}
This trend is consistent with that found in the Fokker-Planck and N-body simulations of multimass systems (see \citealt{2004KiLe} and \citealt{2013HoKi})
and has also been found to emerge at the end of the violent relaxation phase in simulations of the early dynamics of multimass, rotating clusters (\citealt{2021LiVe}).
\setcitestyle{notesep={, }}

In order to further illustrate the internal dynamics driving the observed differences between the rotational properties of stars in different mass groups, we plot the time evolution of the number of stars in each mass group and model in Fig. \ref{fig:mloss}, as well as time evolution of the sum of the $z$ (the $z$-axis coincides with the rotation axis in our simulations) component of angular momentum divided by the total mass for stars with different masses and at different radial distances in Fig. \ref{fig:lzfrac}. The radial cuts for Fig. \ref{fig:lzfrac} are made at the cylindrical half-mass radius for each mass group, and for each mass group the ratio of the group's total angular momentum to the group's total mass is normalised to the initial value of this ratio for the same radial and mass group.

Fig. \ref{fig:lzfrac} illustrates some fundamental aspects of the dynamics of angular momentum evolution and transfer; specifically this figure shows that: 1) in conjunction with Fig. \ref{fig:mloss}, the escape of stars and its dependence on the stellar mass has a significant effect on the angular momentum loss in each system and mass group; 2) angular momentum is transferred from the inner regions to the outer regions; 3) massive stars in each region are transferring part of their angular momentum to the low-mass stars; 4) the angular momentum transfer between massive and low-mass stars is more rapid in the cluster's inner regions.
Although the study of the early dynamical stability and evolution of these systems is not the focus of our investigation, we point out that in our simulations we find that the most rapidly rotating model studied here (the \KRHigh\ model) is characterised by the presence of a bar between $0.4-0.6 t_{\rm rh}$. The growth and evolution of a bar in this model has been discussed in detail by \cite{2013HoKi} and, as discussed in that study, the bar may affect the angular momentum transfer between the different regions of the cluster during the cluster's early ($t < 0.6 t_{\rm rh}$) evolutionary stages.

As a final note, we point out that the three models considered here differ not only in the strength of their initial rotation but also in their density profile; the increase in mass loss, therefore, cannot be ascribed solely to rotation but must rather be due to a combination of structural and kinematic differences.

\subsection{Evolution of velocity anisotropy profiles}
In Fig. \ref{fig:anisotropy}, we show the velocity anisotropy of our models by plotting the time evolution of the radial profile of the ratio of the radial velocity dispersion, $\sigma_{\rm R}$, to the tangential velocity dispersion, $\sigma_{\rm \phi}$, both being evaluated on the plane perpendicular to the rotation axis. 
All systems start with an velocity anisotropy profile characterised by an isotropic core and a halo with increasing radial anisotropy at larger distances from the cluster's centre. Similarly to what has been found in previous studies for non-rotating systems (see e.g. \citealt{2016aTiVe}), our results show that the rotating systems we studied slowly evolve towards an isotropic velocity distribution; this evolution is particularly evident in the outermost regions where the preferential loss of stars on radial/highly-eccentric orbits leads to an isotropic/tangentially anisotropic velocity distribution. 
It is interesting to notice that the anisotropy profiles of the \KRMed\ and \KRHigh\ models gradually develop a dependence on the stellar mass: more massive stars are characterised by a stronger radial anisotropy. This effect is driven by mass loss due to two-body relaxation as shown for each model in Fig. \ref{fig:mloss}: the preferential loss of low-mass stars on radial orbits leads to a more rapid evolution towards an isotropic distribution for low-mass stars, leaving the intermediate and outer regions with mass-dependent velocity anisotropy. 

\subsection{Evolution towards energy equipartition}
In this section, we explore the evolution towards energy equipartition and its dependence on the strength of the cluster's rotation.

In our analysis, we have evaluated the strength of the evolution towards energy equipartition by using the equipartition mass, $m_{\rm eq}$, originally defined in \cite{meq}. The equipartition mass is defined as follows:

\begin{equation}
\resizebox{0.41\textwidth}{!}{$
    \sigma(m) =\left\{\begin{array}{cc}
    \sigma(m_{\rm eq}) \left(\frac{m}{m_{\rm eq}}\right)^{-1/2} &  \rm{if} \ m> m_{\rm eq}\ \ \rm{and} \ m_{\rm eq}>0 \\
    \sigma_0 \exp\left(-\frac{1}{2} \frac{m}{m_{\rm eq}}\right) & \rm{otherwise.}
         \end{array}  \right.$}
    \label{eq:meq}
\end{equation}

As a cluster evolves towards energy equipartition, $m_{\rm eq}$ evolves towards smaller (positive) values while larger values of $m_{\rm eq}$ correspond to systems with a weaker dependence of the velocity dispersion on mass and thus farther from energy equipartition.

Note that this definition allows for negative values of $m_{\rm eq}$. Negative values of $m_{\rm eq}$ describe the case in which the velocity dispersion increases with the stellar mass, a behavior opposite to that expected for systems evolving towards energy equipartition. 
As discussed later, we find that negative values of $m_{\rm eq}$ are needed for a proper fit in the outer regions of some of our models (see \citealt{2021PaVe}, \citeyear{2022PaVe},  where negative values of $m_{\rm eq}$ have been found in the outer regions of isotropic non-rotating systems).

In order to properly fit these functions in systems with bulk rotation, we modify the likelihood function used by \cite{meq} to account for the bulk velocity by subtracting the mean velocity for each star's mass and radial bin, $\bar v(m_{i},R_{i})$, from each star's velocity. The likelihood function is defined as:

\begin{equation}
\resizebox{0.41\textwidth}{!}{$\mathcal{L}=\displaystyle \prod_{i=1}^N\frac{1}{\sqrt{2\pi \sigma^2(m_{i})}}\exp\left[ \frac{-(v_{i}-(\overline v(m_{i},R_{i}))^2}{2\sigma^2(m_{i})} \right]$}
\end{equation}

\noindent where $m_i$ and $v_i$ are the mass and velocity of each star, respectively.
We calculate the equipartition mass, separately, for the three components of the velocity in cylindrical coordinates: $v_{\rm R}$, $v_{\rm \phi}$, and $v_{\rm z}$  and refer to them as $m_{\rm eq, R}$, $m_{\rm eq, \phi}$, and $m_{\rm eq, z}$. 

We  visualise the radial profiles of these three values of $m_{\rm eq}$ in Fig. \ref{fig:meq_inv}, and in Fig. \ref{fig:meq_time} we plot the time evolution of $m_{\rm eq, R}$, $m_{\rm eq, \phi}$, and $m_{\rm eq, z}$ in different cylindrical shells.
These figures show that, for all models, the evolution towards energy equipartition is more rapid in the inner regions which are characterised by a shorter local relaxation time. In the inner regions, $m_{\rm eq}$ is independent of the component of the velocity dispersion; whereas, the evolution of $m_{\rm eq}$ in the intermediate and outer regions proceeds at a different rate for the various velocity dispersion components, indicating that the evolution towards energy equipartition is not isotropic (see also \citealt{2021PaVe} for evidence of anisotropic energy equipartition in non-rotating systems). 
It is interesting to point out, in particular, the more rapid evolution of $m_{\rm eq}$ in the tangential directions (the direction associated with each cluster's internal rotation) for the two more rapidly rotating models. In the outermost regions of the \KRLow\ model, we find that the $z$ component of the velocity dispersion is characterised by a negative $m_{\rm eq}$ (see left panel of Fig. \ref{fig:meq_inv}) corresponding to an anomalous "inverted" energy equipartition similar to that found and discussed in \cite{2021PaVe}.

In order to further understand the evolution towards energy equipartition across the cluster, we have built a 2D map of $m_{\rm eq}$ on the $R$-$z$ meridional plane. As discussed above, $m_{\rm eq}$ quantifies the strength of the dependence of the velocity dispersion on the stellar mass, and its calculation requires a sample of stars large enough to capture this trend. Calculating $m_{\rm eq}$ on a 2D map requires, compared to the 1D profiles shown in Fig. \ref{fig:meq_inv}, an additional sub-sampling of stars in 2D cells which will necessarily increase the noise in the estimate of $m_{\rm eq}$. 
In order to create a spatial map of $m_{\rm eq}$, we created (using the \textsc{R} package \textsc{spatstat}, \citealt{spatstatBook}) a \textit{Dirichlet} tiling based on random subsets of stars in the cluster in $R$-$|z|$ space and, for each tile, we calculated $m_{\rm eq}$ using all of the stars falling in that tile.
We have repeated this procedure multiple times to create different random realizations of the tiling and determined $m_{\rm eq}$ for each tile in each realization. The values of $m_{\rm eq}$ associated with the median $R$ and $|z|$ coordinates within each tile are then smoothed over the $R$-$|z|$ plane using Gaussian kernels. The resulting maps are shown in Fig. \ref{fig:meqmap}.

These maps shed further light on the evolution towards energy equipartition in rotating systems, the spatial anisotropy of this evolution, and its dependence on the velocity components. 
Fig. \ref{fig:meqmap} shows that $m_{\rm eq}$ depends on both $R$ and $z$ and its distribution on the meridional plane is characterised by an increasingly flatter morphology as the cluster's rotation increases.
The anisotropy in the effects of two-body relaxation and their relationship with the cluster morphology are further explored in Fig. \ref{fig:LDMEQmap}. The various panels in this figure show the 2D meridional plane maps of the cluster 3D mass density, the mean mass, the quantity $\sigma^3/(\langle m\rangle \rho)$, which is related to the cluster relaxation time (see e.g. \citealt{2008BiTr}), and of the equipartition mass (here calculated using all of the velocity dispersion components). The maps of the mean mass and the equipartition mass clearly show how the processes of mass segregation and energy equipartition are characterised by an anisotropic evolution correlated with the cluster spatial morphology and the spatial variation of the relaxation time. 
Anisotropic mass segregation has been recently discussed by \cite{2019SzMe}, \cite{2021TiVa}, and \cite{2021LiVe}; here we have further expanded those studies, shown the general anisotropy in the effects of two-body relaxation (mass segregation, evolution towards energy equipartition), and put it in the context of the spatial variation of the cluster structure and relaxation time.

\section{Conclusions}
\label{Conclusions}
We have investigated, by means of N-body simulations, the long-term dynamical evolution of multimass rotating star clusters in a tidal field. We have focused our attention on the effects of two-body relaxation on the cluster’s internal rotation, anisotropy in the velocity distribution, and evolution towards energy equipartition.
The results of our simulations show evidence of a number of kinematic features resulting from the interplay between rotation, stellar escape, two-body relaxation, internal angular momentum transfer, and angular momentum loss due to escaping stars.
Our results can be summarised as follows.
\begin{itemize}
  \item From the study of the time evolution of the rotation curves, we found that as the cluster evolves, the cluster rotational velocity develops a dependence on the stellar mass. Massive stars collectively rotate more rapidly than low-mass stars and the peak of their rotation curves are at a smaller distance from the cluster's centre than for low-mass stars.
    \item We explored the time evolution of the angular momentum for different stellar masses at different radial distances from the cluster's centre. Our results show that angular momentum is transferred from high-mass to low-mass stars and from the inner to the outer regions. As found in previous studies, the strength of rotation in our systems also gradually decreases as a result of the loss of angular momentum carried away by escaping stars.
  \item The systems we studied are characterised by a radially anisotropic velocity distribution. The preferential loss of low-mass stars on radial/high-eccentricity orbits results in the development of a stellar mass dependence of the velocity anisotropy with more massive stars characterised by a more radially anisotropic velocity distribution than low-mass stars.
\item We studied the evolution towards (partial) energy equipartition and found that in the cluster's intermediate and outer regions the degree of energy equipartition depends on the velocity dispersion component; for clusters with a stronger initial rotation, the evolution towards energy equipartition is more rapid in the direction of the rotational velocity.
\item We explored the spatial variation of the mean mass and of the degree of energy equipartition on the $R$-$|z|$ meridional plane. Clusters follow the standard evolution towards mass segregation and energy equipartition, but we find that both spatial segregation and energy equipartition are characterised by an anisotropic flattened spatial variation; this variation is correlated with the flattening in the cluster's structural properties due to the cluster's internal rotation and provides evidence of anisotropy in the effects of two-body relaxation in rotating star clusters.
\end{itemize}

Many observational studies are providing new insights into  the internal kinematics of star clusters. HST and Gaia proper motion studies along with large radial-velocity surveys have led to a significant progress in the characterization of a number of fundamental kinematic properties (such as internal rotation, anisotropy in the velocity distribution, energy equipartition) and their internal radial variation with the distance from the cluster's centre (see e.g. \citealt{2017BeBi}, \citealt{2018KaHu}, \citealt{2018FeLa}, \citealt{2018Biva}, \citealt{2018LiBe}). The results presented in this paper provide new elements for the theoretical framework necessary to interpret current and future observational studies and shed light on the link between the clusters' present-day kinematic properties and their dynamical history.

\section*{ACKNOWLEDGEMENTS}
ALV acknowledges support from a UKRI Future Leaders
Fellowship (MR/S018859/1).

\section*{DATA AVAILABILITY STATEMENT}
The data presented in this article may be shared on reasonable request to the corresponding author.

\bibliographystyle{mnras}
\bibliography{bib}
\end{document}